\documentclass[preprint, 12pt]{elsarticle}
\pdfoutput=1

\usepackage{amssymb}
\usepackage{amsthm}

\usepackage{amsmath}
\usepackage{amssymb}
\usepackage{amsfonts}
\usepackage{pifont}
\usepackage{bbm}
\usepackage{epsfig}
\usepackage{graphics}
\usepackage{color}
\usepackage{multirow}
\usepackage{dsfont}

\renewcommand{\vec}[1]{\mathbf{#1}}

\journal{Computer Physics Communications}

\begin{document}

\begin{frontmatter}



\title{Symbolic computation of the Hartree-Fock energy from a chiral EFT three-nucleon interaction at N$^2$LO}

\author[nscl,msu]{B.~Gebremariam\corref{cor1}}
\ead{gebremar@nscl.msu.edu}
\author[nscl,msu]{S.~K.~Bogner\corref{cor2}}
\ead{bogner@nscl.msu.edu}
\author[nscl,msu,sac]{T.~Duguet}
\ead{thomas.duguet@cea.fr}
\address[nscl]{National Superconducting
Cyclotron Laboratory, 1 Cyclotron Laboratory, East-Lansing, MI
48824, USA}
\address[msu]{Department of Physics and Astronomy,
Michigan State University, East Lansing, MI 48824, USA}
\address[sac]{CEA, Centre de Saclay, IRFU/Service de Physique
Nucl{\'e}aire, F-91191 Gif-sur-Yvette, France}

\begin{abstract}
We present the first of a two-part Mathematica notebook collection that implements a symbolic approach for the
application of the density matrix expansion (DME) to the Hartree-Fock (HF) energy
from a chiral effective field theory (EFT) three-nucleon interaction at N$^2$LO. The final output from the notebooks is a Skyrme-like energy density
functional that provides a quasi-local approximation to the nonlocal HF energy. In this paper,
we discuss the derivation of the HF energy and its simplification in terms of the scalar/vector-isoscalar/isovector parts of the one-body density matrix. Furthermore, a set of steps is described and illustrated on how to extend the approach to other three-nucleon interactions.
\end{abstract}

\end{frontmatter}



{\em PROGRAM SUMMARY}

\begin{small}
\noindent
{\em Program Title:} SymbHFNNN    \\
{\em Catalogue identifier:}                \\
{\em Licensing provisions:}                \\
{\em Programming language:}  Mathematica 7.1\\
{\em Computer:} Any computer running Mathematica 6.0 and later versions\\
{\em Operating system: Windows Xp, Linux/Unix} \\
{\em RAM:} 256 Mb \\
{\em Classification:} 5, 17.16, 17.22\\
{\em Nature of problem:} The calculation of the HF energy from the chiral EFT three-nucleon interaction at N$^2$LO involves tremendous spin-isospin algebra. The problem is compounded by the need to eventually obtain a quasi-local approximation to the HF energy, which requires
the HF energy to be expressed in terms of scalar/vector-isoscalar/isovector parts of the one-body density matrix. The Mathematica notebooks discussed in this paper solve the latter issue.\\
{\em Solution method:} The HF energy from the chiral EFT three-nucleon interaction at N$^2$LO is cast into a form suitable for
an automatic simplification of the spin-isospin traces. Several Mathematica functions and symbolic manipulation techniques are used to obtain the result in terms of the scalar/vector-isoscalar/isovector parts of the one-body density matrix.\\
{\em Running time:} Several hours.
\end{small}




\section{Introduction}\label{Intro}
The nuclear Energy Density Functional (EDF), due to its computational advantages, is the only tractable approach to the calculation of
ground- and excited-state properties of medium to heavy mass nuclei~\cite{bender03}.
Currently, nuclear EDF calculations rely on phenomenologically adjusted Skyrme and Gogny functionals.
Even though these empirical functionals have been successful in providing a satisfactory description of bulk and certain spectroscopic properties of known nuclei, they lack predictive power away from the region where they are fit. In light of this, various groups are pursuing different strategies to improve the predictive power of such phenomenological functionals ~\cite{lesinski06a,Lesinski:2007zz,Margueron:2007uf,Niksic:2008vp,carlsson09,Goriely:2009zz}. One of these approaches is based on the explicit addition of microscopic long-range physics ~\cite{Bogner,Gebremariam2} through the use of the density matrix expansion method~\cite{Negele, Gebremariam1}.

Recent progress in evolving chiral effective field theory (EFT) interactions ~\cite{N3LO, N3LOEGM} to lower resolution scale using the renormalization group methods ~\cite{Vlowk,smooth,Born} makes the construction of microscopically-based energy density functionals a feasible endeavor. Towards such a goal, we derive the Hartree-Fock contribution from chiral EFT two- and three-nucleon interaction at N$^2$LO. Furthermore, the highly nonlocal nature of the resulting energy density functional requires the application of the density matrix expansion~\cite{Negele, Gebremariam1} to derive a Skyrme-like quasi-local energy density functional ~\cite{Bogner, Gebremariam2}. Still, the EDF thus obtained only contains the Hartree-Fock physics such that further correlations must be added to produce any reasonable description of nuclei. In the short term, such an addition of correlations can be done empirically~\cite{stoitsov09a}. Eventually though, it is our goal to compute higher-orders in perturbation theory and design a generalized density matrix expansion that is suited to simplify the non-localities in both space and time that arise beyond the Hartree-Fock level~\cite{rotival09a}.

The tremendous algebra required to derive a Skyrme-like energy density functional from the exact HF energy through the DME makes a manual calculation impractical.
This is especially true for the contribution from the three-nucleon interaction.
In addition, such a derivation displays several features that make it amenable to symbolic automation ~\cite{Wolfram}: (i) it involves many similar and repetitive algebraic steps
(ii) most of it does not involve numerical computation, and  (iii) the part of it that seems to require numerical computation,
such as multidimensional integrals, can be performed using a combination of analytic expansion and symbolic integration~\cite{Gebremariam3}.

In this paper, we present the first part of the Mathematica solution
in which we implemented the calculation of the HF energy from the chiral EFT three-nucleon interaction at N$^2$LO and its subsequent simplification and re-expression in terms of scalar/vector-isoscalar/isovector parts of the one-body density matrix. The second part, which is the subject of a subsequent publication, will deal with the application of the DME to the
output of the first part, thereby yielding a Skyrme-like energy density functional. The paper is structured as follows:
in section \ref{3NFHF}, the starting expression of the HF contribution from the chiral EFT three-nucleon interaction at N$^2$LO is written in such a way that its Mathematica implementation becomes transparent. This is followed by section \ref{Mathe-Implem} where the Mathematica implementation is discussed in detail. The subject of section \ref{moreonpack} is devoted to the organization of the notebooks and the generation of the
Mathematica code itself from the attached Python script, as well as to the modifications required to adapt the application of the notebook to more general three-nucleon interactions and to relax the isospin symmetry imposed in the present work. The last section contains the conclusions.


\section{HF energy from the chiral EFT three-nucleon interaction at N$^2$LO}\label{3NFHF}

\subsection{The $\chi$-EFT three-nucleon interaction at N$^2$LO}
\begin{figure}[hbtp]
  \includegraphics*[width=1.00\columnwidth]
   {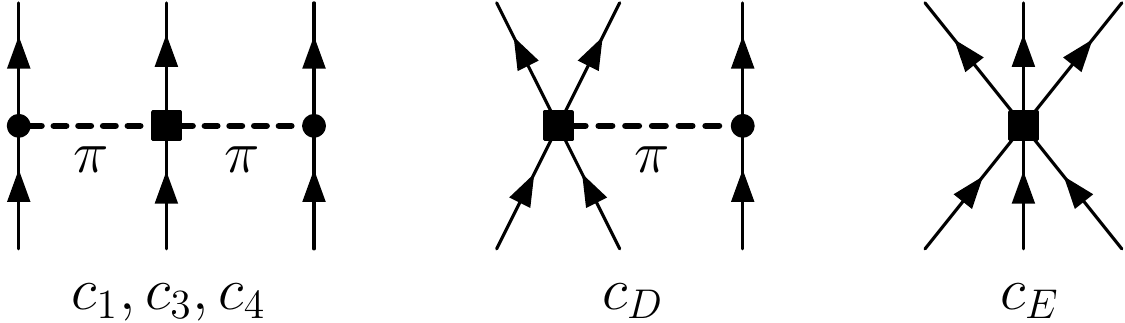}
  \caption{
The two-pion exchange, one-pion exchange, and contact parts of chiral EFT three-nucleon interaction at N$^2$LO and the relevant coupling constants of the interaction.
  \label{fig_all}}
\end{figure}

From a general standpoint, three-nucleon interactions can be written as
\begin{equation}
\langle \vec{k}_1 \vec{k}_2 \vec{k}_3 \lvert \hat{V}_{3N} \rvert
\vec{k}^{\, \prime}_1 \vec{k}^{\, \prime}_2 \vec{k}^{\, \prime}_3 \rangle
\,=\, \frac{1}{\Omega^2} \, \delta_{\vec{k}_1 +
 \vec{k}_2 +   \vec{k}_3, - \vec{k}^{\, \prime}_1 - \vec{k}^{\, \prime}_2 -\vec{k}^{\, \prime}_3}
 \,\hat{V}_{3N}(\vec{k}_1 \vec{k}_2 \vec{k}_3| \vec{k}^{\, \prime}_1  \vec{k}^{\, \prime}_2
 \vec{k}^{\, \prime}_3)\,,
\end{equation}
where $\Omega$ is the volume used in the box-normalization of the momentum basis states  $\lvert \vec{k} \rangle$, $\delta_{\vec{k}_1 +
 \vec{k}_2 +   \vec{k}_3, - \vec{k}^{\, \prime}_1 - \vec{k}^{\, \prime}_2 -\vec{k}^{\, \prime}_3}$
 is the Kronecker delta and
$\hat{V}_{3N}(\vec{k}_1 \vec{k}_2 \vec{k}_3| \vec{k}^{\, \prime}_1  \vec{k}^{\, \prime}_2
 \vec{k}^{\, \prime}_3)$ is a matrix element in momentum space and an operator in
spin-isospin space. In the above expression, the dependence of
the interaction on spin and isospin degrees of freedom is not displayed.
The chiral EFT three-nucleon interaction at N$^2$LO has three main
components (i) the E-term (ii) the D-term and (iii) the C-term that are diagrammatically represented in Fig.~\ref{fig_all}.
As the present work relies only on the operator structure of the interaction which is provided below, we do not specify the actual values of the
various constants that appear in the interaction. Refer to Ref.~\cite{epelbaum} for details.
Furthermore, we neglect the regulator function as we are dealing with a HF calculation. This is justified because the HF calculation does not probe high-momentum single-particle states. Specifically, one can neglect the regulator
so long as $k_F \ll \Lambda$, where $k_F$ is the local Fermi momentum and $\Lambda$ is the cutoff parameter employed in
the interaction. The absence of the regulator function makes the interaction local in coordinate space. Even though
section ~\ref{extension} illustrates how to extend the approach to non-local interactions, the locality of the interaction
brings significant reduction in the complexity of the second part of the problem, viz, the application of the DME to produce a quasi-local EDF.

\subsection{Remarks on the notation}
The following definitions hold for the various operators that appear in the analytical expressions for E-, D- and C-terms of the interaction
that are stated below. $\vec{\sigma}_i \equiv (\sigma_{i,x},\,\sigma_{i,y},\,\sigma_{i,z} )$ and
$\vec{\tau}_i \equiv (\tau_{i,x},\,\tau_{i,y},\,\sigma_{i,z} )$ refer to the spin and isospin Pauli matrices of the $i^{th}$ particle
respectively. The momentum transfer for the $i^{th}$ particle is denoted with $\vec{q}_i \equiv \vec{k}_i - \vec{k}^{\prime}_i$ where
$\vec{k}_i$ and $\vec{k}^{\prime}_i$ are the out- and in-momentum coordinates of the particle. The scalar product of two vectors $\vec{a}$ and
$\vec{b}$ is denoted with $\vec{a} \cdot \vec{b}$. Furthermore, repeated indices imply summation over those indices (Einstein convention).
\subsubsection{The E-term}
The E-term, which is a three-nucleon contact interaction,
is the simplest part of the interaction. The corresponding expression reads
\begin{equation}\label{Epart}
\hat{V}_E (\vec{k}_1 \vec{k}_2 \vec{k}_3| \vec{k}^{\, \prime}_1
\vec{k}^{\, \prime}_2
 \vec{k}^{\, \prime}_3) \,\equiv\, E \bigl(\, \vec{\tau}_1 \cdot \vec{\tau}_2 +
 \vec{\tau}_2 \cdot \vec{\tau}_3  + \vec{\tau}_3 \cdot \vec{\tau}_1
 \,\bigr)\,,
\end{equation}
where
\begin{equation}
E \, \equiv \, \frac{C_E}{f^{4}_\pi \Lambda_{\chi}}\,.
\end{equation}


\subsubsection{The D-term}
The D-term involves a one-pion exchange and a contact interaction. Its analytical form reads
\begin{eqnarray}\label{Dpart}
\hat{V}_D (\vec{k}_1 \vec{k}_2 \vec{k}_3| \vec{k}^{\, \prime}_1
\vec{k}^{\, \prime}_2
 \vec{k}^{\, \prime}_3) \,&\equiv& \,- \frac{g_A}{4 f^2_\pi} \frac{C_D}{f^2_{\pi} \Lambda_{\chi}}
\,\biggl( \frac{\sigma_1 \cdot \vec{q}_2 \, \sigma_2 \cdot
\vec{q}_2}{q^2_2 + m^2_\pi} \tau_1 \cdot \tau_2 \, + \,
\frac{\sigma_2 \cdot \vec{q}_3 \, \sigma_3 \cdot \vec{q}_3}{q^2_3 +
m^2_\pi} \nonumber\\
&& \, \times \,\tau_2 \cdot \tau_3  \, + \, \frac{\sigma_3 \cdot \vec{q}_1
\, \sigma_1 \cdot \vec{q}_1}{q^2_1 + m^2_\pi} \tau_3 \cdot \tau_1
\biggr)\,.
\end{eqnarray}


\subsubsection{The C-term}
The C-term involves two-pion exchanges and its analytic form is
\begin{eqnarray}\label{Cpart}
\hat{V}_C (\vec{k}_1 \vec{k}_2 \vec{k}_3| \vec{k}^{\, \prime}_1
\vec{k}^{\, \prime}_2
 \vec{k}^{\, \prime}_3) \,&\equiv &\,\biggl(\frac{g_A}{2 f_{\pi}}\biggr)^2
\,\biggl(\frac{\sigma_1 \cdot \vec{q}_1 \, \sigma_2 \cdot
\vec{q}_2}{( q^2_1 + m^2_\pi)(q^2_2 + m^2_\pi)} F^{\alpha
\beta}_{123} \tau^{\alpha}_1 \tau^{\beta}_2 + \frac{\sigma_2 \cdot
\vec{q}_2 }{ q^2_2 + m^2_\pi} \nonumber \\
&& \hspace{-.4in}\times \,\frac{ \sigma_3 \cdot \vec{q}_3}{q^2_3 +
m^2_\pi} \, F^{\alpha \beta}_{231} \tau^{\alpha}_2 \tau^{\beta}_3
+\frac{\sigma_3 \cdot \vec{q}_3 \, \sigma_1 \cdot \vec{q}_1}{( q^2_3
+ m^2_\pi)(q^2_1 + m^2_\pi)} F^{\alpha \beta}_{312} \tau^{\alpha}_3
\tau^{\beta}_1
 \biggr)\,,
\end{eqnarray}
where
\begin{equation}\label{Falphabeta}
F^{\alpha \beta}_{i j k}\, \equiv  \, \delta_{\alpha \beta}\,\bigl[\,
-4\,\frac{c_1 m^2_\pi}{f^2_\pi} \, + \,2\,
\frac{c_3}{f^2_\pi}\vec{q}_i \cdot \vec{q}_j \, \bigr] \, + \,
\frac{c_4}{f^2_\pi} \epsilon^{\alpha \beta \gamma} \tau^{\gamma}_{k}
\, \vec{\sigma}_k \cdot (\vec{q}_i \times \vec{q}_j)\,.
\end{equation}


\subsection{Basic form of the HF energy from the $\chi$-EFT three-nucleon interaction at N$^2$LO}\label{basicEqn3NFHF}
The three-nucleon interaction can in general be decomposed as a sum of three terms
\begin{equation}\label{3NFsplitting}
\hat{V}_{3N} \,\equiv\, \hat{V}_{ 1 2 } + \hat{V}_{ 2 3 } +\hat{V}_{ 1 3 } \,,
\end{equation}
where $\hat{V}_{i j}$ is symmetric in $i$ and $j$. Specifically, for the
chiral EFT three-nucleon interaction at N$^2$LO, $V_{ij}$ depends on $\vec{q}_i$ and
$\vec{q}_j$ and, in general, on the spin-isospin coordinates of the three nucleons. After a few basic algebraic manipulations, the
HF energy from the three-nucleon interaction can be expressed in terms of only one of the three $\hat{V}_{ij}$
operators, e.g. $\hat{V}_{23}$, as
\begin{eqnarray}\label{Simplified3NFHF}
\langle \hat{V}^{\text{HF}}_{3 N} \rangle &=&\, \frac{1}{2} \,
\sum_{i j k} \, \langle i  j  k \lvert \hat{V}_{23} (1
 -2 \, P_{13} - P_{23} + 2 \,P_{12} P_{23} ) \lvert i j k\rangle\,,
\end{eqnarray}
where $P_{lm}$ denotes the exchange operator (of particles $l$ and
$m$) whereas  $i,j \, \text{and}\, k$ denote occupied HF single-particle states. Note that for ease of notation,
we are using the single-particle basis that diagonalizes the one-body density matrix of the HF Slater determinant.
$P_{lm}$ is defined as
\begin{equation}
P_{lm} \, \equiv \, P^{r}_{lm} \,P^{\sigma}_{lm}\,P^{\tau}_{lm}\,,
\end{equation}
with $P^{r}_{lm}$ the coordinate exchange operator and  the spin-isospin exchange operators given by
$P^{\sigma}_{lm} \equiv ( 1 + \vec{\sigma}_l \cdot \vec{\sigma}_m )/2$ and $P^{\tau}_{lm} \equiv ( 1 + \vec{\tau}_l \cdot \vec{\tau}_m )/2$.
One can identify three groups of terms in Eq. \eqref{Simplified3NFHF}: direct, single exchange and double exchange
terms\footnote{This should not be confused with one- and two-pion exchanges contribution to the three-nucleon interaction.}.
The direct term corresponds to the expectation value of $\hat{V}_{23}$, the single-exchange
term to the expectation value of $\hat{V}_{23} (-2 P_{13} - P_{23})$ and the double-exchange term
to that of $2\, \hat{V}_{23}  P_{12} P_{23}$

\begin{eqnarray}
\langle V^{\text{HF},\text{dir}}_{3N} \rangle \,& \equiv &\, \frac{1}{2} \,
\sum_{i j k} \, \langle i  j  k \lvert \hat{V}_{23}  \lvert i j k\rangle\,,\label{3NFHFbasisdir}\\
\langle V^{\text{HF},\text{1x}}_{3N} \rangle \,& \equiv &\, \frac{1}{2} \,
\sum_{i j k} \, \langle i  j  k \lvert \hat{V}_{23} (-2 P_{13} - P_{23})  \lvert i j k\rangle\,,\label{3NFHFbasis1x}\\
\langle V^{\text{HF},\text{2x}}_{3N} \rangle \,& \equiv & \,
\sum_{i j k} \, \langle i  j  k \lvert \hat{V}_{23}  P_{12} P_{23}  \lvert i j k\rangle\,\label{3NFHFbasis2x}.
\end{eqnarray}

As the derivation of the a Skyrme-like quasi-local EDF from the exact HF energy requires the application of the DME,
we need to express the HF energy in the $\lvert \vec{r} \rangle
\otimes \lvert \sigma \rangle \otimes \lvert \tau \rangle$ single-particle basis. This is due to the fact that the DME, as formulated ~\cite{Negele, Gebremariam1},
is most naturally applicable to the one-body density matrix expressed in coordinate space. Hence, we
perform inverse-Fourier transformation of the interaction to express the matrix elements in
$\lvert \vec{r} \rangle \otimes \lvert \sigma
\rangle \otimes \lvert \tau \rangle$ single-particle basis. This transformation leaves
the spin-isospin dependencies untouched. Confining the
discussion to the $\vec{k}$ dependence, we have

\begin{eqnarray}\label{Vinrspace}
\langle \vec{r}_1 \vec{r}_2 \vec{r}_3 \lvert \hat{V}_{23}\rvert
\vec{r}^{\,\prime}_1 \vec{r}^{\,\prime}_2
\vec{r}^{\,\prime}_3\rangle\, &=&\, \delta(\vec{r}_1 -
\vec{r}^{\,\prime}_1)\, \delta(\vec{r}_2 - \vec{r}^{\,\prime}_2)\,
\delta(\vec{r}_3 - \vec{r}^{\,\prime}_3) \,\nonumber\\
&&\hspace{.75in}\null\times\mathbb{V}_{2 3}
(\vec{r}_2-\vec{r}_1, \vec{r}_3 -\vec{r}_1)\,,
\end{eqnarray}
where
\begin{equation}\label{V23}
\mathbb{V}_{2 3} (\vec{r}_2-\vec{r}_1, \vec{r}_3-\vec{r}_1)\equiv\frac{1}{(2 \pi)^6}
 \,\int  d \vec{q}_2  d \vec{q}_3 \,
e^{i\vec{q}_2.(\vec{r}_2 -\vec{r}_1)} \, e^{i\vec{q}_3.(\vec{r}_3 -\vec{r}_1)}\,V_{2 3} (\vec{q}_2, \vec{q}_3)\,.
\end{equation}
At this point, we do not actually perform the integrals over the momentum coordinates
\footnote{Except for the E-term of the interaction which is a trivial three-nucleon contact
interaction, thereby yielding simple delta functions as shown in Eq. \eqref{sampleHFdir}.} in Eq \eqref{V23}.
Rather, Eq. \eqref{V23} is used
 used as it is, resulting in five-dimensional integrals in Eqs. \eqref{3NFHFdir}-\eqref{3NFHF2x} that are performed after the application of the DME as discussed in Ref. ~\cite{Gebremariam2}.

The next target is to rewrite Eqs. \eqref{3NFHFbasisdir}-\eqref{3NFHFbasis2x} in a form transparent for Mathematica
implementation (in $\lvert \vec{r} \rangle \otimes \lvert \sigma
\rangle \otimes \lvert \tau \rangle$ single-particle basis). We illustrate the steps required to achieve that with
Eq. \eqref{3NFHFbasis2x}, for which we have
\begin{eqnarray}
\langle V^{\text{HF},\text{2x}}_{3N} \rangle \,& \equiv & \,
\sum_{i j k} \, \langle i  j  k \lvert \hat{V}_{23}  P_{12} P_{23}  \lvert i j k\rangle\, \nonumber\\
&=&\, \sum_{i j k} \sum_{\sigma^{\prime}_1 .. \sigma_3} \sum_{\tau^{\prime}_1 .. \tau_3} \int \,\prod^{3}_{m=1}d \vec{r}^{\prime}_m
\,\prod^{3}_{n=1} d \vec{r}^{\prime}_n  \, \langle i  j  k \lvert
\vec{r}{\prime}_1 \sigma^{\prime}_1 \tau^{\prime}_1 \,
\vec{r}{\prime}_2 \sigma^{\prime}_2 \tau^{\prime}_2 \,
\vec{r}{\prime}_3 \sigma^{\prime}_3 \tau^{\prime}_3 \rangle \nonumber\\
&&\, \times \,\langle
\vec{r}{\prime}_1 \sigma^{\prime}_1 \tau^{\prime}_1 \,
\vec{r}{\prime}_2 \sigma^{\prime}_2 \tau^{\prime}_2 \,
\vec{r}{\prime}_3 \sigma^{\prime}_3 \tau^{\prime}_3 \lvert
\hat{V}_{23}  P^{\sigma \tau}_{12} P^{\sigma \tau}_{23}  \lvert
\vec{r}_1 \sigma_1 \tau_1 \,
\vec{r}_2 \sigma_2 \tau_2 \,
\vec{r}_3 \sigma_3 \tau_3 \rangle \nonumber\\
&& \, \times \, \langle
\vec{r}_1 \sigma_1 \tau_1 \,
\vec{r}_2 \sigma_2 \tau_2 \,
\vec{r}_3 \sigma_3 \tau_3 \lvert
P^{\vec{r}}_{12} P^{\vec{r}}_{23} \lvert
i j k\rangle\,,\label{basicstepforMathe}
\end{eqnarray}
where we used completeness relations
\begin{equation}
\sum_{\sigma_1\, .. \,\sigma_3} \sum_{\tau_1\, .. \,\tau_3} \int  \,\prod^{3}_{i=1} d \vec{r}_i\,\lvert \vec{r}_1 \sigma_1 \tau_1 \,
\vec{r}_2 \sigma_2 \tau_2 \,
\vec{r}_3 \sigma_3 \tau_3 \rangle \, \langle
 \vec{r}_1 \sigma_1 \tau_1 \,
\vec{r}_2 \sigma_2 \tau_2 \,
\vec{r}_3 \sigma_3 \tau_3 \lvert \, = \,  \mathds{1}\,,
\end{equation}
and $P^{\sigma \tau}_{lm} \equiv P^{\sigma}_{lm} P^{\tau}_{lm}$. We split the particle exchange operator such that the coordinate part
acts on the wave-functions while the spin-isospin piece is taken care of along with the interaction.
Let $\vec{X}_i$ represent $(\vec{r}_i\sigma_i \tau_i)$ such that the one-body density matrix reads as
\begin{eqnarray}
\varrho (\vec{X}_j, \vec{X}_k)\,\equiv\,
\varrho (\vec{r}_j \sigma_j \tau_j , \vec{r}_k \sigma_k \tau_k ) \,
\equiv
\sum_{i} \, \varphi^{\ast}_{i}(\vec{r}_k \, \sigma_k \,\tau_k)
\, \varphi_{i}(\vec{r}_j \,\sigma_j\, \tau_j) \,,
\end{eqnarray}
where the sums is over occupied single-particle HF states. Making use of this, we define another quantity, which we call the auxiliary density matrix, as
\begin{eqnarray}
\varrho^{i} (\vec{X}_j, \vec{X}_k)\,\equiv\,
\varrho\, (\vec{r}_j \sigma^{\prime}_i \tau^{\prime}_i , \vec{r}_k \sigma_i \tau_i ) \,, \label{auxiliaryrho}
\end{eqnarray}
where $i \,\epsilon\, \{1,\,2,\,3 \}$. Basically, the spin-isospin coordinates of this quantity are those of the $i^{th}$ particle.
Applying the steps demonstrated in Eq. \eqref{basicstepforMathe} and using Eqs. \eqref{auxiliaryrho}, \eqref{Vinrspace}-\eqref{V23}, one can express the
direct, single-exchange and double-exchange parts of the three-nucleon interaction HF
energy as
\begin{eqnarray}
\langle V^{\text{HF},\text{dir}}_{3N} \rangle \,& = &\, \frac{1}{2} \,\text{Tr}_1
\text{Tr}_2\text{Tr}_3 \biggl[  \int d \vec{r}_1 d \vec{r}_2 d \vec{r}_3 \,
\varrho^{1}(\vec{X}_1) \,\varrho^{2}(\vec{X}_2) \,\varrho^{3}(\vec{X}_3) \,\nonumber\\
&& \, \quad\quad\quad\quad\quad\quad\quad \times \, \mathbb{V}_{2 3} (\vec{r}_2-\vec{r}_1, \vec{r}_3-\vec{r}_1) \biggr]\,, \label{3NFHFdir}\\
\langle V^{\text{HF},\text{1x}}_{3N} \rangle \,& = &\,-
\,\text{Tr}_1 \text{Tr}_2\text{Tr}_3 \biggl[ \int d \vec{r}_1 d \vec{r}_2 d
\vec{r}_3 \, \varrho^{1}(\vec{X}_3, \vec{X}_1)
\,\varrho^{2}(\vec{X}_2) \,\varrho^{3}(\vec{X}_1,\vec{X}_3) \nonumber\\
&&\, \quad\quad\quad\quad\quad\quad\quad \times \mathbb{V}_{2 3} (\vec{r}_2-\vec{r}_1, \vec{r}_3-\vec{r}_1) \, P^{\sigma \tau}_{13}\biggr]\,\nonumber\\
 &&
\,- \,\frac{1}{2}\,\text{Tr}_1
\text{Tr}_2\text{Tr}_3  \biggl[ \int d \vec{r}_1 d \vec{r}_2 d \vec{r}_3 \,
\varrho^{1}(\vec{X}_1) \,\varrho^{2}(\vec{X}_3,\vec{X}_2) \,\varrho^{3}(\vec{X}_2,\vec{X}_3)\nonumber\\
&&\, \quad\quad\quad\quad\quad\quad\quad \times\mathbb{V}_{2 3} (\vec{r}_2-\vec{r}_1, \vec{r}_3-\vec{r}_1) \,
  P^{\sigma \tau}_{23} \biggr] \,,\label{3NFHF1x}\\
 \langle V^{\text{HF},\text{2x}}_{3N} \rangle \,& = & \,\text{Tr}_1
\text{Tr}_2\text{Tr}_3  \int d \vec{r}_1 d \vec{r}_2 d \vec{r}_3 \,\biggl[
\varrho^{1}(\vec{X}_2,\vec{X}_1) \,\varrho^{2}(\vec{X}_3,\vec{X}_2) \,\varrho^{3}(\vec{X}_1,\vec{X}_3) \,
\nonumber\\
&&\, \quad\quad\quad\quad\quad\quad\quad \times\mathbb{V}_{2 3} (\vec{r}_2-\vec{r}_1, \vec{r}_3-\vec{r}_1)\, P^{\sigma \tau}_{23} P^{\sigma \tau}_{12} \biggr]\,\label{3NFHF2x}\,,
 \end{eqnarray}
where $\varrho^{i}(\vec{X}_j) \equiv \varrho^{i}(\vec{X}_j,\vec{X}_j)$ and $\text{Tr}_i$ refers to tracing
over the spin and isospin coordinates of the $i^{\text{th}}$ particle.
The key to understand the form of these equations is the splitting of
the particle exchange operator, performed in Eqs. \eqref{basicstepforMathe}, that results in
the spin-isospin coordinates of each particle to be grouped in a single auxiliary density matrix.
These are the basic equations to be implemented directly in Mathematica.
The next section shows that the implementation of these equations is transparent, which would not have been the case without the trick used to group the spin-isospin coordinates
of each particle in a single auxiliary density matrix.

\section{Mathematica Implementation}\label{Mathe-Implem}
Starting from Eqs. \eqref{3NFHFdir}-\eqref{3NFHF2x},
the implementation consists of the following two components: (i) the automated tracing operation of spin-isospin matrices and
(ii) the re-expression of the HF energy in terms of the scalar/vector-isoscalar/isovector parts of the one-body density matrix.

\subsection{Automated Tracing}\label{AutomatedTracing}
The first task to be automated is the tracing operations in Eqs. \eqref{3NFHFdir}-\eqref{3NFHF2x}. This requires
expressing the auxiliary density matrix in terms of its scalar/vector-isoscalar/isovector parts.
We adopt the notation used in Ref. ~\cite{Perli} where for two vectors, $\vec{v}$ and $\vec{w}$ in isospin space,
$\vec{v}\circ\vec{w}$ denotes their scalar product. Hence

\begin{eqnarray}
\varrho^i (\vec{r}^{\,\prime} \sigma^{\prime}_i \tau^{\prime}_i, \vec{r} \sigma_i \tau_i)\, &=&
\, \frac{1}{4} \biggl( \rho^{i}_0 (\vec{r}^{\,\prime} , \vec{r} )\, \delta_{\sigma^{\prime}_i \sigma_i } \,\delta_{\tau^{\prime}_i \tau_i } \,
+ \, \vec{\rho}^{\,i}_1 (\vec{r}^{\, \prime} , \vec{r} )   \,\circ\, \vec{\tau}_{ \tau^{\prime}_i \tau_i}\,\delta_{ \sigma^{\prime}_i \sigma_i} \nonumber\\&&
+ \,\vec{s}^{\,i}_0 (\vec{r}^{\,\prime},\vec{r}  )\cdot \vec{\sigma}_{\sigma^{\prime}_i \sigma_i}
\,\delta_{\tau^{\prime}_i \tau_i } + \, \vec{s}^{\,i}_1 (\vec{r}^{\,\prime} , \vec{r})\cdot
\vec{\sigma}_{ \sigma^{\prime}_i \sigma_i} \, \circ \,\vec{\tau}_{\tau^{\prime}_i \tau_i } \biggr)\,.
\end{eqnarray}
with the scalar-isoscalar, scalar-isovector, vector-isoscalar and vector-isovector components are given, respectively, as
\begin{eqnarray}
\rho^{i}_0 (\vec{r}^{\,\prime} , \vec{r} ) \, &\equiv&\,\sum_{\sigma \tau}\, \varrho^i (\vec{r}^{\,\prime} \sigma_i \tau_i, \vec{r} \sigma_i \tau_i) \, ,\label{scasca} \\
\vec{\rho}^{i}_1 (\vec{r}^{\,\prime} , \vec{r} ) \, &\equiv&\, \sum_{\sigma \tau^{\prime} \tau }\, \varrho^i (\vec{r}^{\,\prime} \sigma_i \tau^{\prime}_i, \vec{r} \sigma_i \tau_i) \,\tau_{\tau_i \tau^{\prime}_i} \, , \label{scavec} \\
\vec{s}^{i}_0 (\vec{r}^{\,\prime} , \vec{r} ) \, &\equiv&\,\sum_{\sigma^{\prime} \sigma \tau }\, \varrho^i (\vec{r}^{\,\prime} \sigma^{\prime}_i \tau_i, \vec{r} \sigma_i \tau_i) \, \sigma_{\sigma_i \sigma^{\prime}_i} \,, \label{vecsca}  \\
\vec{s}^{i}_1 (\vec{r}^{\,\prime} , \vec{r} ) \, &\equiv&\, \sum_{\sigma^{\prime} \sigma \tau^{\prime} \tau } \, \varrho^i (\vec{r}^{\,\prime} \sigma^{\prime}_i \tau^{\prime}_i, \vec{r} \sigma_i \tau_i) \,\sigma_{\sigma_i \sigma^{\prime}_i} \,\tau_{\tau_i \tau^{\prime}_i}  \,, \label{vecvec}
\end{eqnarray}
where $\sigma_{\sigma_i \sigma^{\prime}_i} \equiv \langle \sigma_i \lvert \sigma \rvert \sigma^{\prime}_i \rangle $
and $\tau_{\tau_i \tau^{\prime}_i} \equiv \langle \tau_i \lvert \tau \rvert \tau^{\prime}_i \rangle $, with the operators $\vec{\sigma} \equiv
(\sigma_x,\, \sigma_y,\,\sigma_z)$ and $\vec{\tau} \equiv (\tau_x, \tau_y, \tau_z)$.
In contrast to the one-body density matrix, we, occasionally, refer to the scalar/vector-isoscalar/isovector components of the one-body density matrix as nonlocal densities. Note that we maintain the explicit reference to the index $i$ as it will be useful in the Mathematica representation of the respective quantity. At this point, we identify all the basic quantities that need to be
represented by their own Mathematica symbol. These are listed in table \ref{symboltable}.
In this table, the index $i$ plays
the following two roles:
\begin{itemize}
\item[(i)] It identifies the coordinate dependence of the scalar/vector-isoscalar/isovector nonlocal densities. For instance, for the double-exchange part of the HF energy (as given in Eq.\eqref{3NFHF2x}), the three auxiliary density matrices have the coordinate dependence
\begin{eqnarray}
\varrho^1 (\vec{X}_j, \vec{X}_k) \, &=&\,\varrho^1 (\vec{X}_2, \vec{X}_1)\,, \nonumber\\
\varrho^2 (\vec{X}_j, \vec{X}_k) \, &=&\,\varrho^2 (\vec{X}_3, \vec{X}_2)\,, \nonumber\\
\varrho^3 (\vec{X}_j, \vec{X}_k) \, &=&\,\varrho^3 (\vec{X}_1, \vec{X}_3)\,. \nonumber \,,
\end{eqnarray}
implying that the scalar/vector-isoscalar/isovector nonlocal densities derived from these will have their coordinate dependencies fixed accordingly.
For instance, the Mathematica symbol $\rho10$ refers to the scalar-isoscalar nonlocal density extracted from the first auxiliary density matrix.
According to Eqs.\eqref{3NFHFdir}-\eqref{3NFHF2x}, this entails that $\rho10$ denotes the non/local density $\rho_0 (\vec{r}_1)$ for the case of direct contribution, $\rho_0 (\vec{r}_3,\vec{r}_1)$ for the first part of the single-exchange contribution, $\rho_0 (\vec{r}_1) $ for the second-part of the single-exchange contribution and
$\rho_0(\vec{r}_2,\vec{r}_1)$ for the double-exchange contribution.
\item[(ii)] One should be convinced that for the spin and isospin traces, the index $i$  plays the role of particle-id. Hence, the spin and isospin matrices obtained from the auxiliary density matrices are to be grouped with the respective matrices obtained from the interaction and spin-isospin exchange operators.
\end{itemize}
\begin{table}[ht]
\caption{List of basic quantities and their corresponding symbols.} 
\centering 
\begin{tabular}{c c c} 
\hline\hline 
Quantity & Symbol & Remark \\ [0.5ex] 
\hline 
\\
\multirow{2}{*}{$\vec{\sigma}$}   &  \multirow{2}{*}{$\sigma$}   & The vector containing the three Pauli
matrices:\\&& $\sigma = \{\sigma_x , \sigma_y , \sigma_z \}$. \\
\\
\multirow{2}{*}{$\vec{\tau}$}  &   \multirow{2}{*}{$\tau$} &  The vector containing the three Pauli
matrices:\\&& $\tau = \{\tau_x , \tau_y , \tau_z \}$.\\
\\

\multirow{6}{*}{$\vec{s}_{i j}$}  & \multirow{6}{*}{$sij$} & The vector containing the components
of the vector\\&& density: $\vec{s} = \{s_x,\, s_y,\, s_z \}.$
The index $i\, \epsilon\, \{1,2,3\} $ denotes\\&& the first, second or
third density in Eqs.\eqref{3NFHFdir}-\eqref{3NFHF2x} while\\&& $j\, \epsilon \,\{0,1\}$ refers to
the isoscalar and isovector components\\&& respectively. For example, the Mathematica symbol $s10$ \\&& defines
the vector $s10=\{s10x,\, s10y,\, s10z\}$. \\
\\

\multirow{5}{*}{$\rho_{ij}$}  & \multirow{5}{*}{$\rho ij$}  &  The scalar part of the density matrix.
The index \\&& $i\, \epsilon\, \{1, \,2,\,3\} $
denotes the first, second or third \\&& density in Eqs.\eqref{3NFHFdir}-\eqref{3NFHF2x} while $j\,
\epsilon \,\{0,\,1\}$ refers to\\&& the isoscalar and isovector parts, respectively.\\&&
For example, the Mathematica symbol $\rho11$ \\&& defines
the vector $\rho11=\{\rho11x,\, \rho11y,\, \rho11z\}$.\\
\\

\multirow{3}{*}{$q_i$}   & \multirow{3}{*}{$qi$}  &  The momentum vector $q$ vector in the interaction \\&& where $i \, \epsilon \, \{2,3\}$ labels the particle.
For example,\\&& the symbol $q2$ defines the vector $q2 =\{q2x, \, q2y, \,q2z\}$.
\\
\\ [1ex] 
\hline 
\end{tabular}
\label{symboltable} 
\end{table}

Once the symbols of the various quantities at play are identified, the next step is to define
the composite symbols, viz, vectors and tensors. In the following we list such definitions.
\begin{align}
\text{In[1]:}&=iM = \text{IdentityMatrix}[2]; \nonumber\\
\text{In[2]:}&=\sigma x = \{\{0,\,1\},\,\{1,\,0\}\}; \nonumber\\
\text{In[3]:}&=\sigma y = \{\{0,\,-i\},\,\{i,\,0\}\}; \nonumber\\
\text{In[4]:}&=\sigma z = \{\{1,\,0\},\,\{0,\,-1\}\} ;\nonumber\\
\text{In[5]:}&=\tau x = \{\{0,\,1\},\,\{1,\,0\}\} ;\nonumber\\
\text{In[6]:}&=\tau y = \{\{0,\,-i\},\,\{i,\,0\}\} ;\nonumber\\
\text{In[7]:}&=\tau z = \{\{1,\,0\},\,\{0,\,-1\}\} ;\nonumber\\
\text{In[8]:}&=\sigma = \{\sigma x,\, \sigma y,\, \sigma z\} ;\nonumber\\
\text{In[9]:}&=\tau = \{\tau x,\, \tau y,\, \tau z\}; \nonumber\\
\text{In[10]:}&= q2 = \{q2x,\, q2y,\, q2z\} ;\nonumber\\
\text{In[11]:}&= q3 = \{q3x,\, q3y,\, q3z\} ;\nonumber\\
\text{In[12]:}&= \rho 11 = \{\rho 11x,\, \rho 11y,\, \rho 11z\} ;\nonumber\\
\text{In[13]:}&= s10 = \{s10x,\, s10y,\, s10z\} ;\nonumber\\
\text{In[14]:}&= s11 = \{\{s11xx,\, s11xy,\, s11xz\},\, \{s11yx,\, s11yy,\, s11yz\}, \,\nonumber\\
		&\quad\quad\quad\quad\quad \{s11zx,\, s11zy,\, s11zz\}\} ;\nonumber\\
\text{In[15]:}&= \rho 21 = \{\rho 21x,\, \rho 21y,\, \rho 21z\} ;\nonumber\\
\text{In[16]:}&= s20 = \{s20x,\, s20y,\, s20z\} ;\nonumber\\
\text{In[17]:}&= s21 = \{\{s21xx,\, s21xy,\, s21xz\},\, \{s21yx,\, s21yy,\, s21yz\}, \,\nonumber\\
		&\quad\quad\quad\quad\quad \{s21zx,\, s21zy,\, s21zz\}\} ;\nonumber\\
\text{In[18]:}&= \rho 31 = \{\rho 31x,\, \rho 31y,\, \rho 31z\} ;\nonumber\\
\text{In[19]:}&= s30 = \{s30x,\, s30y,\, s30z\} ;\nonumber\\
\text{In[20]:}&= s31 = \{\{s31xx,\, s31xy,\, s31xz\},\, \{s31yx,\, s31yy,\, s31yz\}, \,\nonumber\\
		&\quad\quad\quad\quad\quad \{s31zx,\, s31zy,\, s31zz\}\} ;\nonumber
\end{align}

The actual tracing operation involves: (i) grouping together all
spin-isospin operators that originate from the interaction, spin-isospin exchange operators or an
auxiliary density matrix carrying the same particle-id. This should be done without mixing their actual ordering as only
matrices with different particle-ids commute with each other. (ii) Performing the spin and isospin traces separately. In the
following, we demonstrate the steps followed using the simplest term from the HF energy of chiral EFT
three-nucleon interaction (at N$^2$LO), viz, the direct part of the HF energy from the E-part of the
interaction. We denote this quantity by $\langle V^{\text{HF},E,\text{dir}}_{3N}\rangle$  and, according to
Eqs. \eqref{Epart} and \eqref{3NFHFdir}, it reads

\begin{eqnarray}\label{sampleHFdir}
\langle V^{\text{HF},E,\text{dir}}_{3N} \rangle \,&=& \,
\frac{1}{2}\,E\, \text{Tr}_1 \text{Tr}_2\text{Tr}_3  \int d
\vec{r}_1 d \vec{r}_2 d \vec{r}_3 \, \varrho^{1}(\vec{X}_1)
\,\varrho^{2}(\vec{X}_2) \,\varrho^{3}(\vec{X}_3)\nonumber\\
&&\quad\quad\quad\quad \times\,  \delta (\vec{r}_1 -\vec{r}_2)\, \delta (\vec{r}_1 -\vec{r}_3) \,\tau_2 \cdot
\tau_3\,.
\end{eqnarray}
The Mathematica code to perform the automated tracing reads
\begin{eqnarray}\label{examplecode1}
\text{In[19]:}&=&\text{V3NDirectHFE}=\biggl[\text{  }\sum _{\text{$\alpha $1}=1}^3 \frac{1}{128}\nonumber\\&&
\biggl(\text{$\rho $10}*\text{Tr}[\text{iM}]*\text{Tr}[\text{iM}] \nonumber\\&&
\quad\quad\quad\quad\quad+\text{Tr}[\text{iM}]*\text{Tr}[(\text{$\rho $11}.\tau )]+\text{Tr}[(\text{s10}.\sigma )]*\text{Tr}[\text{iM}] \nonumber\\ &&
\quad\quad\quad\quad\quad+\,\sum _{\text{$\beta $1}=1}^3\sum _{\text{$\beta $2}=1}^3\text{s11}[[\text{$\beta $1}]][[\text{$\beta $2}]]\text{Tr}[\sigma [[\text{$\beta $1}]]]*\text{Tr}[\tau [[\text{$\beta $2}]]]\biggr) \nonumber\\ &&
\biggl(\text{$\rho $20}*\text{Tr}[\text{iM}]*\text{Tr}[\text{iM}.\tau [[\text{$\alpha $1}]]]+\text{Tr}[\text{iM}]*\text{Tr}[(\text{$\rho $21}.\tau ).\tau [[\text{$\alpha $1}]]]\nonumber\\ &&
\quad\quad\quad\quad\quad+ \text{Tr}[(\text{s20}.\sigma )]*\text{Tr}[\text{iM}.\tau [[\text{$\alpha $1}]]]\nonumber\\&&
\quad\quad\quad\quad\quad+\sum _{\text{$\beta $3}=1}^3\sum _{\text{$\beta $4}=1}^3\text{s21}[[\text{$\beta $3}]][[\text{$\beta $4}]]\text{Tr}[\sigma [[\text{$\beta $3}]]]*\text{Tr}[\tau [[\text{$\beta $4}]].\tau [[\text{$\alpha $1}]]]\biggr)\nonumber\\ &&
\biggl(\text{$\rho $30}*\text{Tr}[\text{iM}]*\text{Tr}[\text{iM}.\tau [[\text{$\alpha $1}]]]
+\text{Tr}[\text{iM}]*\text{Tr}[(\text{$\rho $31}.\tau ).\tau [[\text{$\alpha $1}]]]\nonumber\\ &&
\quad\quad\quad\quad\quad+ \text{Tr}[(\text{s30}.\sigma )]*\text{Tr}[\text{iM}.\tau [[\text{$\alpha $1}]]] \nonumber\\&&
\quad\quad\quad\quad\quad+ \sum _{\text{$\beta $5}=1}^3\sum _{\text{$\beta $6}=1}^3\text{s31}[[\text{$\beta $5}]][[\text{$\beta $6}]]\text{Tr}[\sigma [[\text{$\beta $5}]]]*\text{Tr}[\tau [[\text{$\beta $6}]].\tau [[\text{$\alpha $1}]]]\biggr),\,\nonumber\\ &&
\text{Assumptions}\to \text{AssumptionsList}\biggr]\,,
\end{eqnarray}
where the Mathematica symbol AssumptionsList contains
\begin{eqnarray}\label{assumptionslist}
\text{In[19]:}&=&\text{AssumptionsList} \,=\,\{\text{$\rho $10}\text{==}\text{$\rho $20}\text{==}\text{$\rho $30},\,\,\,\text{$\rho $11z}\text{==}\text{$\rho $21z}\text{==}\text{$\rho $31z},\,\nonumber\\ &&
\text{$\rho $11y}\text{==}\text{$\rho $21y}\text{==}\text{$\rho $31y}==0,\,\,\,
\text{$\rho $11x}\text{==}\text{$\rho $21x}\text{==}\text{$\rho $31x}==0,\,\nonumber\\ &&
\text{s10x}\text{==}\text{s20x}\text{==}\text{s30x},\,\,\,
\text{s10y}\text{==}\text{s20y}\text{==}\text{s30y},\,\nonumber\\ &&
\text{s10z}\text{==}\text{s20z}\text{==}\text{s30z},\,\,\,
\text{s11xx}\text{==}\text{s21xx}\text{==}\text{s31xx}==0,\,\nonumber\\ &&
\text{s11xy}\text{==}\text{s21xy}\text{==}\text{s31xy}==0,\,\,\,
\text{s11xz}\text{==}\text{s21xz}\text{==}\text{s31xz},\,\nonumber\\ &&
\text{s11yx}\text{==}\text{s21yx}\text{==}\text{s31yx}==0,\,\,\,
\text{s11yy}\text{==}\text{s21yy}\text{==}\text{s31yy}==0,\,\nonumber\\ &&
\text{s11yz}\text{==}\text{s21yz}\text{==}\text{s31yz},\,\,\,\,
\text{s11zx}\text{==}\text{s21zx}\text{==}\text{s31zx}==0,\,\nonumber\\ &&
\text{s11zy}\text{==}\text{s21zy}\text{==}\text{s31zy}==0,\,\,\,
\text{s11zz}\text{==}\text{s21zz}\text{==}\text{s31zz}\}\,.
\end{eqnarray}

While the tracing operation has been explained, the list enclosed in $\text{Assumptions}\to$ warrants a few lines of explanation.
The list contains two groups (i) a set of statements in which symbols are equated with each other. These are solely the symbols referring to
various scalar/vector-isoscalar/isovector densities having the same coordinate dependence, which can be traced back to the occurrence of the two delta functions in Eq. \eqref{sampleHFdir}. Note that this is due to the fact that the E-term is a three-nucleon contact interaction. For the D- and C-terms, a more complicated coordinate dependence occurs that results in there being no symbols equated in the respective AssumptionsList. (ii) The second group contains a list of symbols that are
set to zero. It is based on the fact that the HF energy is isospin invariant, as is the starting interaction.
Moreover, the present application forbids proton-neutron mixing, implying that single-particle wave-functions are eigenstates of the isospin-projection operator on the $z$ axis. Hence, we enforce proton-neutron symmetry by performing rotation
in isospin space Ref.~\cite{Perli} using the conserved symmetry operator
\begin{equation}\label{pnsymmetry}
\hat{U}_{pn} \, = \, i \, \text{exp}\biggl( - \frac{i}{2} \pi  \hat{\tau}_3 \biggr)\,,
\end{equation}
which aligns the quantities along the $z$ axis in isospin space. Such a step yields a tremendous reduction in the size of the resulting
expressions. Thus, in the final expression of the HF energy, only isoscalar and the $z$ component of the isovector densities appear.

Even though Eq. \eqref{examplecode1} corresponds to the simplest part of the problem, its Mathematica code requires
a difficult mix of proper ordering of the operators and correct labeling of the dummy indices
for the summation operations. This becomes truly involved when dealing, for instance, with
the double-exchange part of the HF energy from
the C-term of the interaction. To address this problem, we have written a Python
script that can be used to generate the required Mathematica code. Refer to section \ref{moreonpack}
for details.

At this stage, the output from Mathematica is such that the
various vectors and tensors occur in terms of their components which makes the expression too lengthy
and impractical. To illustrate this particular point, we choose a small part of the single-exchange contribution from the D-term.
Refer to section \ref{moreonpack} for details on the organization of the notebooks.
Even in this case, the output is much simpler than the typical outputs from the double-exchange of the D- and any of the C-terms\footnote{The output of the automated tracing from the double-exchange of the D-term is tens of pages long while that of the single- and double-exchange of the the C-term
are much longer than that.}.
Keeping that in mind, the output reads
{\allowdisplaybreaks[4]
\begin{align}
\text{Out[19]:}& =
\frac{1}{8} \text{$\rho $10}\, \biggl(\, 2 \text{q3y} \text{q3z}\, (-3 \text{s20z} \text{s30y}\,-\,3 \text{s20y} \text{s30z}\,
+\,\text{s21zz} \text{s31yz}\,\nonumber\\&
+\,\text{s21yz} \text{s31zz})\,+\,2 \text{q3x} (\text{q3y} (-3 \text{s20y} \text{s30x}\,-\,
3 \text{s20x} \text{s30y}\,\nonumber\\&
+\,\text{s21yz} \text{s31xz}\,+\,\text{s21xz} \text{s31yz})\,+\,\text{q3z} (-3 \text{s20z} \text{s30x}\,
-\,3 \text{s20x} \text{s30z}\,\nonumber\\&
+\,\text{s21zz} \text{s31xz}\,+\,\text{s21xz} \text{s31zz}))\,+\,\text{q3x}^2 (-3 \text{s20x} \text{s30x}\,
+\,3 \text{s20y} \text{s30y}\,\nonumber\\&
+\,3 \text{s20z} \text{s30z}\,+\,\text{s21xz} \text{s31xz}\,-\,\text{s21yz} \text{s31yz}\,
-\,\text{s21zz} \text{s31zz}\,-\,3 \text{$\rho $20} \text{$\rho $30}\,\nonumber\\&
+\,\text{$\rho $21z} \text{$\rho $31z})\,+\,\text{q3y}^2 (3 \text{s20x} \text{s30x}\,-\,3 \text{s20y} \text{s30y}\,
+\,3 \text{s20z} \text{s30z}\,\nonumber\\&
-\,\text{s21xz} \text{s31xz}\,+\,\text{s21yz} \text{s31yz}\,-\,\text{s21zz} \text{s31zz}\,-\,3 \text{$\rho $20} \text{$\rho $30}\,
+\,\text{$\rho $21z} \text{$\rho $31z})\,\nonumber\\&
+\,\text{q3z}^2 (3 \text{s20x} \text{s30x}\,+\,3 \text{s20y} \text{s30y}\,
-\,3 \text{s20z} \text{s30z}\,-\,\text{s21xz} \text{s31xz}\,\nonumber\\&
-\,\text{s21yz} \text{s31yz}\,
+\,\text{s21zz} \text{s31zz}\,
-\,3 \text{$\rho $20} \text{$\rho $30}\,+\,\text{$\rho $21z} \text{$\rho $31z})\biggr)\nonumber\\
&
\frac{1}{8} \text{$\rho $11}\, \biggl(\, 2 \text{q3y} \text{q3z}\, (-3 \text{s21xz} \text{s30y}\,-\,3 \text{s21yz} \text{s30z}\,
+\,\text{s20zz} \text{s31yz}\,\nonumber\\&
+\,\text{s20y} \text{s31zz})\,+\,2 \text{q3x} (\text{q3y} (-3 \text{s21yz} \text{s30x}\,-\,
3 \text{s21xz} \text{s30y}\,\nonumber\\&
+\,\text{s20y} \text{s31xz}\,+\,\text{s20x} \text{s31yz})\,+\,\text{q3z} (-3 \text{s21xz} \text{s30x}\,
-\,3 \text{s21xz} \text{s30z}\,\nonumber\\&
+\,\text{s20z} \text{s31xz}\,+\,\text{s20z} \text{s31zz}))\, +  ... \,\biggr)\,.
\end{align}
}
Hence, the manual simplification of such an expressions is not feasible. The next section describes the technique we
used to automate the task of writing these expressions in terms of the scalar/vector-isoscalar/isovector\footnote{From here onwards, we
use the term isovector freely even though we have set the first and second components to zero (only the third is nonzero). Refer to the discussion immediately following Eq. \eqref{assumptionslist}.} parts of the density matrix: $\{ \rho_0, \, \rho_1,\, \vec{s}_0,\,\vec{s}_1\}$.

\subsection{Scalar/vector-isoscalar/isovector re-expression}\label{ScaIsoVecIsovecdecomposition}
The main objective of this part of the code is to rewrite the output of the automated parsing step
in terms of the scalar/vector-isoscalar/isovector parts of the one-body density matrix with the proper coordinate dependence. The approach consists of two basic steps. Initially we form the eight groups that arise from choosing the scalar or vector components of for each of the three density matrices that are multiplied, which we represent schematically as
\begin{equation}
\text{groups} = \{\{0|1\} \otimes \{0|1\}  \otimes \{0|1\}  \},
\end{equation}
where $\{0|1\}$
represents taking one of the scalars $\{\rho_0, \, \rho_1\}$ or one of the vectors $\{ \vec{s}_0,\,\vec{s}_1\}$ of that density matrix. Once a particular group is formed, the next step involves forming the list of all scalars and vectors including
momentum vectors coming from the interaction.

In the following, a complete illustration is given for the double-exchange of the D-term. The list of scalars
and vectors that we have for instance, for the $\{011\}$ group of the double-exchange of the D-term,
are
\begin{equation}
\text{group011} =\{\vec{q}_3,\,\vec{q}_3,\,\rho^1_0,\,\rho^1_1, \,\vec{s}^2_0, \,\vec{s}^2_1,\,\vec{s}^3_0,
\,\vec{s}^3_1, \}\,,
\end{equation}
where the superscripts \{1,\,2,\,3\} specify the id of the original density matrix. Note that, for
each density matrix, we have two possible choices (isoscalar/isovector) even after the scalar/vector choice has been made. In this set, the only momentum vector that appears is $q_3$ and it appears twice as can be seen from the combination of Eqs. \eqref{Dpart} and \eqref{3NFHF2x}.

Since energy is invariant with respect to rotation in coordinate, spin and isospin spaces, we next form the list of all possible scalars from
the selected group. In each scalar, we can pick only one of the two possible choices originating from the same
density matrix. For instance, referring to $\text{group011}$, we cannot have $\rho^1_0$ and $\rho^1_1$ in the same scalar. Hence, the scalars that can be formed from $\text{group011}$ are
\begin{eqnarray}\label{scalarlist}
\text{scalars011} &=& \{\, \rho^1_0 \,(\vec{q}_3.\vec{q}_3)\, (\vec{s}^2_0.\vec{s}^3_0), \,
 \rho^1_0 \,(\vec{q}_3.\vec{q}_3)\, (\vec{s}^2_1.\vec{s}^3_1), \,
\rho^1_1 \,(\vec{q}_3.\vec{q}_3)\, (\vec{s}^2_0.\vec{s}^3_1), \,\nonumber\\&&
\rho^1_1 \,(\vec{q}_3.\vec{q}_3)\, (\vec{s}^2_1.\vec{s}^3_0), \,
\rho^1_0 \,(\vec{q}_3.\vec{s}^2_0)\, (\vec{q}_3.\vec{s}^3_0), \,
\rho^1_0 \,(\vec{q}_3.\vec{s}^2_1)\, (\vec{q}_3.\vec{s}^3_1), \,\nonumber\\&&
\rho^1_1 \,(\vec{q}_3.\vec{s}^2_0)\, (\vec{q}_3.\vec{s}^3_1), \,
\rho^1_1 \,(\vec{q}_3.\vec{s}^2_1)\, (\vec{q}_3.\vec{s}^3_0) \,\}.
\end{eqnarray}

The subsequent step involves using Mathematica's $\text{SolveAlways}$ function to obtain the coefficients
for each term of $\text{scalars011}$. To complete the discussion, let us explicit the part of the Mathematica code that
implements the key ideas of this section
\begin{align}
\text{in[19]:} & s11z = \{s11xz, \, s11yz, \, s11zz\}; \nonumber\\
\text{in[19]:} & s21z = \{s21xz, \,s21yz, \,s21zz\};\nonumber \\
\text{in[19]:}& s31z = \{s31xz,\, s31yz, \,s31zz\};\nonumber\\
\text{in[19]:}& \text{V3NDoubleExchangeHFDgroup011} = \text{Simplify}[\text{V3NDoubleExchangeHFD}, \,\nonumber\\
&\text{Assumptions} \to \{s10x==0,\,s10y==0,\,s10z==0,\,\nonumber\\
& s11xz==0,\,s11yz==0,\,s11zz==0,\,\nonumber\\
&{\rho}20==0,\, {\rho}21z==0, \,{\rho}30==0,{\rho}31z==0\}]\,,
\end{align}
where V3NDoubleExchangeHFD is the symbol for the double-exchange contribution from the D-term
obtained after the automated tracing. In the actual implementation, the sheer size of the V3NDoubleExchangeHFD forces one to break the expression further into subcomponents. Refer to section \ref{moreonpack} for details. In the above code,
the definitions for $\{s11xz,\,s11yz,\,s11zz\}$ allows one to collect the nonzero parts of the respective tensors, viz, the $z$ components in isospin space as discussed in section \ref{AutomatedTracing}, in to a vector. The code in $\text{Assumptions}\to$ block
simply sets the parts of the density matrices that are zero
in the chosen group, viz, $group011$. For details, refer to the submitted Mathematica code. Once we obtain
the particular group of interest, in this case $group011$, the coefficients of the various scalar terms listed in Eq. \eqref{scalarlist}
are calculated using the code

\begin{align}
\text{in[19]:}& \text{CoefficientList011} = \text{SolveAlways}\bigl[
V3NDoubleExchangeHFDgroup011 \nonumber\\& \, -\,
\bigl(a1  \, q3.q3 \, {\rho}10\, s20.s30 \, + \,  a2 \, q3.q3 \,{\rho}10\, s21z.s31z
\,+ \, a3 \, q3.q3 \,{\rho}11z \, s20.s31z \,\nonumber\\&
+\, a4 \, q3.q3\, {\rho}11z\,  s21z.s30 \,
+ \, a5\,{\rho}10 \, q3.s20 \, q3.s30 \,+ \,a6 \,{\rho}10\,q3.s21z\, q3.s31z \, \nonumber\\&
+ \,a7 \,{\rho}11z\, q3.s20 \, q3.s31z \, + \, a8 \,{\rho}11z\,q3.s21z \,q3.s30
\bigr)
==0
,\nonumber\\&
\{\rho10,\,\rho20,\,\rho30,\,\rho11z,\,\rho21z,\,\rho31z,\, q3x,\,q3y,\,q3z,\,s10x,\,s10y,\,\nonumber\\&s10z,\,s20x,\,
s20y,\,s20z,\,s30x,\,s30y,\,s30z,\,s11xz,\,s11yz,\,\nonumber\\&
s11zz,\,s21xz,\,s21yz,\,s21zz,\,s31xz,\,s31yz,\,s31zz\}\bigr]
\end{align}
The output symbol CoefficientList011 will contain the the values of the coefficients $\{a_1, \,a_2,\,....a_8\}$.
In this case

\begin{align}
\text{in[19]:} \text{CoefficientList011}  = &\{a1 \to -3/16, \, a2 \to 1/16,\,a3 \to -3/16,\,\nonumber\\
&\,a4 \to 1/16,\,a5 \to 3/8, \, a6 \to -1/8,\nonumber\\
&\, a7 \to 3/8,\,a8 \to -1/8\}\,.
\end{align}
Finally a part of the code does the automatic replacement of the coordinate dependencies.
As for the example used in the present section, one is dealing with a part of the double-exchange from the D-term, which implies
that the coordinate dependencies of the three density matrices, according to Eq. \eqref{3NFHF2x}, are
\begin{eqnarray}
\varrho^1 (\vec{X}_j, \vec{X}_k) \, &=&\,\varrho^{1}(\vec{X}_2, \vec{X}_1)\,, \nonumber\\
\varrho^2 (\vec{X}_j, \vec{X}_k) \, &=&\,\varrho^{2}(\vec{X}_3, \vec{X}_2)\,, \nonumber\\
\varrho^3 (\vec{X}_j, \vec{X}_k) \, &=&\,\varrho^{3}(\vec{X}_1, \vec{X}_3)\,. \nonumber
\end{eqnarray}
Thus far the multidimensional integrals and the interaction constants that we have in Eqs.\eqref{3NFHFdir}-\eqref{3NFHF2x}
did not take part in any of the symbolic manipulations. Once the  proper coordinate dependencies are
restored, we insert the interaction constants and the multidimensional integrals resulting in the final expression for V3NDoubleExchangeHFDgroup011, denoted as $\langle V^{D2x}_{011}\rangle$,
\begin{align}\label{Dtermdouble011}
\langle V^{D2x}_{011} \rangle \, =& \,
\frac{-g_A  }{4 f^2_\pi}\frac{C_D}{f^2_\pi \Lambda_x} \,
\frac{1}{16} \, \int d \vec{r}_2 d \vec{r}_3 \,\int  \frac{1}{(2
\pi)^3}d \vec{q}_3 \, e^{i \vec{q}_3.(\vec{r}_3 -\vec{r}_2)}
\,\frac{ q^\beta_3  q^\gamma_3}{q^2_3 + m^2_\pi}\,\biggl[
\delta_{\beta \gamma} \delta_{\mu \nu } \nonumber\\
& \biggl( -3\, \rho_0(\vec{r}_2, \vec{r}_1) s^{\mu}_0 (\vec{r}_3, \vec{r}_2) s^{\nu}_0 (\vec{r}_2, \vec{r}_3)
\,+\,
\rho_0(\vec{r}_2, \vec{r}_1) s^{\mu}_1 (\vec{r}_3, \vec{r}_2) s^{\nu}_1 (\vec{r}_2, \vec{r}_3)\,\nonumber\\
&
-3\,\rho_1(\vec{r}_2, \vec{r}_1) s^{\mu}_0 (\vec{r}_3, \vec{r}_2) s^{\nu}_1 (\vec{r}_2, \vec{r}_3)
\,+\,
\rho_1(\vec{r}_2, \vec{r}_1) s^{\mu}_1 (\vec{r}_3, \vec{r}_2) s^{\nu}_0 (\vec{r}_2, \vec{r}_3)\biggr)\nonumber\\
&
\, + \,3\,\rho_0(\vec{r}_2, \vec{r}_1) s^{\beta}_0 (\vec{r}_3, \vec{r}_2) s^{\gamma}_0 (\vec{r}_2, \vec{r}_3)
\,-\,
\rho_0(\vec{r}_2, \vec{r}_1) s^{\beta}_1 (\vec{r}_3, \vec{r}_2) s^{\gamma}_1 (\vec{r}_2, \vec{r}_3) \,\nonumber\\
&
\,+ \, 3\,\rho_1(\vec{r}_2, \vec{r}_1) s^{\beta}_0 (\vec{r}_3, \vec{r}_2) s^{\gamma}_1 (\vec{r}_2, \vec{r}_3)\,-\,
\rho_1(\vec{r}_2, \vec{r}_1) s^{\beta}_1 (\vec{r}_3, \vec{r}_2) s^{\gamma}_0 (\vec{r}_2, \vec{r}_3)
 \biggr].
\end{align}
To find the total value of the contribution from the double-exchange of the D-term,
similar analysis is carried out for the remaining seven groups and their results added to Eq. \eqref{Dtermdouble011}.
Naively counting, one has to calculate about seventy-two groups\footnote{For the three parts of the interaction,
we have direct, single-exchange and double-exchange with each requiring eight groups.} of varying sizes to get
the total contribution of the three-nucleon interaction. This shows the tremendous size of the algebra required to calculate the total HF energy and
hence the need for automation.

\section{More on the Mathematica notebooks}\label{moreonpack}
\subsection{Automatic generation of the Mathematica codes and organization of the notebooks}
As mentioned in section \ref{AutomatedTracing}, writing the Mathematica codes is an error-prone
procedure. This is mainly due to the fact that one has to include a large number of summation indices in a single expression and
maintain proper operator orders. Consequently, we wrote a python script
that can be used to generate the Mathematic codes automatically, given the proper parameters. Refer to the submitted files for details.

The Mathematica implementation of the calculation of the HF energy and re-expression in terms of scalar/vector-isoscalar/isovector parts of the density matrix is carried out in eight notebooks. The contribution from the E- is implemented in a
single notebook and the same for the D-term. The C-term is implemented in the remaining six notebooks. Refer to the submitted code for details.

\section{Possible Extensions}\label{extension}
In this section, we discuss the steps that can be taken to extend the approach so that one can treat non-local interactions
and allow for proton-neutron mixing.

\subsection{Treating non-local interactions}
The approach can be extended to treat nonlocal interactions. The starting Eqs.\eqref{3NFHFdir}-\eqref{3NFHF2x} need to be modified for nonlocal interactions
as
\begin{eqnarray}
\langle V^{\text{HF},\text{dir}}_{3N} \rangle \,& = &\, \frac{1}{2} \,\text{Tr}_1
\text{Tr}_2\text{Tr}_3 \biggl[ \int \prod^{3}_{i=1} d \vec{r}_i \prod^{3}_{j=1} d \vec{r}^{\prime}_i \,
\varrho^{1}(\vec{X}_1, \vec{X}^{\prime}_1) \,\varrho^{2}(\vec{X}_2, \vec{X}^{\prime}_2) \,\varrho^{3}(\vec{X}_3,\vec{X}^{\prime}_3) \,\nonumber\\
&& \, \quad\quad\quad\quad\quad\quad\quad \times \, \langle \vec{r}_1 \vec{r}_2 \vec{r}_3 \, \lvert \mathbb{V}_{2 3} \rvert \, \vec{r}^{\prime}_1 \vec{r}^{\prime}_2 \vec{r}^{\prime}_3 \rangle\, \biggl],
\label{3NFHFdirnonlocal}\\
\langle V^{\text{HF},\text{1x}}_{3N} \rangle \,& = &\,-
\,\text{Tr}_1 \text{Tr}_2\text{Tr}_3 \biggl[ \int d \prod^{3}_{i=1} d \vec{r}_i \prod^{3}_{j=1} d \vec{r}^{\prime}_i
\, \varrho^{1}(\vec{X}_3, \vec{X}^{\prime}_1)
\,\varrho^{2}(\vec{X}_2,\vec{X}^{\prime}_2) \,\varrho^{3}(\vec{X}_1,\vec{X}^{\prime}_3) \nonumber\\
&&\, \quad\quad\quad\quad\quad\quad\quad \times \, \langle \vec{r}_1 \vec{r}_2 \vec{r}_3 \, \lvert \mathbb{V}_{2 3} \rvert \, \vec{r}^{\prime}_1 \vec{r}^{\prime}_2 \vec{r}^{\prime}_3 \rangle\, P^{\sigma \tau}_{13}\,\biggl] \nonumber\\
 &&
\,- \,\frac{1}{2}\,\text{Tr}_1
\text{Tr}_2\text{Tr}_3 \biggl[ \int \prod^{3}_{i=1} d \vec{r}_i \prod^{3}_{j=1} d \vec{r}^{\prime}_i\,
\varrho^{1}(\vec{X}_1, \vec{X}^{\prime}_1) \,\varrho^{2}(\vec{X}_3,\vec{X}^{\prime}_2) \,\varrho^{3}(\vec{X}_2,\vec{X}^{\prime}_3)\nonumber\\
&&\, \quad\quad\quad\quad\quad\quad\quad \times \, \langle \vec{r}_1 \vec{r}_2 \vec{r}_3 \, \lvert \mathbb{V}_{2 3} \rvert \, \vec{r}^{\prime}_1 \vec{r}^{\prime}_2 \vec{r}^{\prime}_3 \rangle\,
  P^{\sigma \tau}_{23}\biggl]\,,\label{3NFHF1xnonlocal}\\
 \langle V^{\text{HF},\text{2x}}_{3N} \rangle \,& = & \,\text{Tr}_1
\text{Tr}_2\text{Tr}_3 \biggl[ \int  \prod^{3}_{i=1} d \vec{r}_i \prod^{3}_{j=1} d \vec{r}^{\prime}_i\,
\varrho^{1}(\vec{X}_2,\vec{X}^{\prime}_1) \,\varrho^{2}(\vec{X}_3,\vec{X}^{\prime}_2) \,\varrho^{3}(\vec{X}_1,\vec{X}^{\prime}_3) \,
\nonumber\\
&&\, \quad\quad\quad\quad\quad\quad\quad \times\, \langle \vec{r}_1 \vec{r}_2 \vec{r}_3 \, \lvert \mathbb{V}_{2 3} \rvert \, \vec{r}^{\prime}_1 \vec{r}^{\prime}_2 \vec{r}^{\prime}_3 \rangle\, P^{\sigma \tau}_{23} P^{\sigma \tau}_{12}\biggl]\,\label{3NFHF2xnonlocal}\,.
 \end{eqnarray}
Starting with this, the same automated spin-isospin tracing, scalar/vector-isoscalar/isovector re-expression can be carried out
as described in secs. \ref{AutomatedTracing} and \ref{ScaIsoVecIsovecdecomposition}. This easy extension remains valid throughout
the first part of the problem, specifically, calculation and re-expression of the HF energy in terms of scalar/vector-isoscalar/isovector
parts of one-body density matrix. This is in contrast to the second part of the problem, application of the DME to produce a Skyrme-like
energy density functional, where the non-locality of the interaction makes the problem significantly harder.

The traditional Skyrme phenomenological interactions ~\cite{skyrme56} employed in nuclear structure calculations are much easier to treat as they
are zero-range non-local interactions. Hence, we demonstrate how the present symbolic approach can be applied directly to calculate
the energy density functional from the three-nucleon part of Skyrme interactions. From the outset, it should be clear that we don't need to further apply the DME to obtain a local energy density functional as
the interaction is of zero-range. The situation is partly similar to the E-term of the realistic chiral EFT three-nucleon interaction. However, the fact that we have gradient operators in the traditional Skyrme phenomenological interactions make the derivation harder than the corresponding derivation for the E-term.

We start with a schematic three-nucleon Skyrme-like interaction
\begin{eqnarray}\label{skyrmelikeinteraction}
\langle \vec{r}_1 \vec{r}_2 \vec{r}_3 \, \lvert \mathbb{V}_{2 3} \rvert \,
 \vec{r}^{\prime}_1 \vec{r}^{\prime}_2 \vec{r}^{\prime}_3 \rangle \,&=& \, \bigl(\,\vec{k}^2_{23} \, + \,\vec{k}^{\prime \, 2}_{23} \,\bigr) \,
\delta (\vec{r}_1-\vec{r}^{\prime}_1)\,\delta (\vec{r}_2-\vec{r}^{\prime}_2)\,\delta (\vec{r}_3-\vec{r}^{\prime}_3)\, \nonumber\\
&& \quad\quad \quad \, \times \, \delta (\vec{r}_1-\vec{r}_2)\,\delta (\vec{r}_2-\vec{r}_3)\,,
\end{eqnarray}
where according to the usual notation of Skyrme interactions $\vec{k}_{ij}= i/2 (\vec{\nabla}_i -\vec{\nabla}_j)$ and
$\vec{k}^{\prime}_{ij}= -i/2 (\vec{\nabla}^{\prime}_i -\vec{\nabla}^{\prime}_j)$. Our convention is that, when using such an interaction
in Eqs. \eqref{3NFHFdirnonlocal}-\eqref{3NFHF2xnonlocal}, gradients must act on the density matrices prior to making use of the delta functions.

The next step in the symbolic simplification is to form the set of rules based on the relationship between the nonlocal
scalar/vector-isoscalar/isovector densities with the various local densities that arise in Skyrme energy density functionals.
A list and properties of these local densities and the relationship
between these local densities and the nonlocal one they originate from can be found in Ref. ~\cite{engel75}. For instance
\begin{eqnarray}
\biggl[\biggr(\vec{\nabla}^2_i + \vec{\nabla}^{\prime 2}_j \biggr) \rho_0 (\vec{r}_i,\vec{r}^{\prime}_j) \biggr] \,
\bigg{|}_{\vec{r}_i =\vec{r}^{\prime}_j =\vec{R}}\,=\,\vec{\nabla}^2 \rho_0(\vec{R}) \, -\, 2 \,\tau_0(\vec{R})\,,
\end{eqnarray}
where $\rho_0(\vec{R})$ is the local scalar-isoscalar density and $\tau_0(\vec{R})$ is the local scalar-isoscalar kinetic density.
These relationships can be expressed as a Mathematica replacement rules involving the symbols representing these quantities.
What follows is a chain of replacement rule operations to obtain the final form of the local energy density functional.
To illustrate the complete set of steps, we have coded and submitted a Mathematica notebook that
derives the local energy density functional resulting from the the direct part (Eq. \eqref{3NFHFdirnonlocal}) of the schematic three-nucleon Skyrme-like interaction given in Eq.\eqref{skyrmelikeinteraction}.  The most important advantage of this automation is that once
the rules for mapping the nonlocal densities onto the local ones are established, calculating the energy density functional
for different spin-isospin and gradient structures requires a negligible fraction of the effort required to perform the corresponding derivation
manually.

\subsection{Proton-neutron mixing}
Allowing for proton-neutron mixing makes possible to produce a more general skyrme-like energy density functional.
The merit of the present approach is that this can be done directly by removing the second group of assumptions
discussed just after Eq. \eqref{assumptionslist}. The main negative side effect that one faces in doing so relates to the size of
the resulting expressions that increase significantly, resulting in a possible failure of Mathematica's SolveAlways method.
Usually, this problem can be alleviated by breaking the unhandled expressions into smaller pieces that can always be
solved by SolveAlways. Moreover, allowing for proton-neutron mixing forces substantial changes when re-expressing the formula
in terms of the scalar/vector-isoscalar/isovector parts of the one-body density matrix. The reason is that one has to use the full
scalar-isovector, $\rho_{ij}$, and vector-isovector, $\vec{s}_{ij}$, symbols instead of their projections along the $z$ axis in isospin space.
As a result, one has to take this into account when constructing all possible scalars as described in section \ref{ScaIsoVecIsovecdecomposition}.
Another approach can be to perform the calculations with a conserved proton-neutron symmetry, as is done in the present work, and
map the resulting expression to the case of proton-neutron mixing.

\section{Comments and conclusions}
The first of a two-part symbolic approach to the derivation of a Skyrme-like quasi-local energy density functional from the HF energy of the chiral EFT three-nucleon interaction at N$^2$LO was discussed and implemented in a collection of Mathematica notebooks. We have demonstrated that the tremendous spin-isospin algebra required to express the HF energy
in terms of the coordinate-space scalar/vector-isoscalar/isovector parts of the one-body density matrix can be effectively tackled by using Mathematica. In the second part of the work, which will be the subject of a subsequent publication, we show that the
further application of the density matrix expansion method needed to generate the Skyrme-like energy density functional
can also be completely automated. Possible future directions of research could be:
(i) as a whole, the current Mathematica implementation conserves proton-neutron symmetry. Hence, one
can allow proton-neutron mixing by relaxing this constraint, thereby arriving at a more general Skyrme-like energy  density functional. The steps that are required to achieve this have been outlined in the present contribution. (ii) We have demonstrated that the approach can be generalized to treat nonlocal interactions such as the quasi-local Skyrme interactions. Calculating the energy density functional starting from a Skyrme-interaction (including two- and three-nucleon parts) with a rich spin-isospin and gradient structures is a tedious task. In light of the recent interest along these lines (see Ref.~\cite{Margueron:2009} and references therein), extending the scheme to treat this problem could be useful. (iii) One can envision expanding the present work in such a way that first-order pairing correlations are treated along with the HF part, viz, performing HFB (Hartree-Fock-Bogoliubov) calculations. Combining this extension with proton-neutron mixing, one can have a starting Skyrme-like functional that can be used to handle proton-neutron pairing correlations as discussed in Ref.~\cite{Perli}. (iv) Implementing a similar scheme to treat four-nucleon interactions can also be one area of extension.

\section{Acknowledgements}
We thank R.~J.~Furnstahl for useful discussions. This work was supported in part by the National
Science Foundation under Grant No. PHY-0758125 and the UNEDF
SciDAC Collaboration (DOE Grant DE-FC02-07ER41457).


\label{}



\end{document}